\documentclass[sigconf,nonacm]{acmart}
\usepackage{amsmath,amsfonts}
\usepackage{algorithmic}
\usepackage{algorithm}
\usepackage{array}
\usepackage{xurl}
\usepackage[caption=false,font=normalsize,labelfont=sf,textfont=sf]{subfig}
\usepackage{textcomp}
\usepackage{stfloats}
\usepackage{url}
\usepackage{verbatim}
\usepackage{graphicx}
\usepackage{listings}
\usepackage{tcolorbox}
\tcbuselibrary{breakable}
\usepackage{booktabs}
\usepackage{multirow}
\usepackage{subcaption}
\usepackage{tabularx}
\newtcolorbox{rqsummary}{
  breakable,
  colback=gray!5,
  colframe=gray!60,
  boxrule=0.5pt,
  arc=2pt,
  left=6pt,
  right=6pt,
  top=6pt,
  bottom=6pt
}
\newcommand{\Code}[1]{%
  {\small\fontsize{9.5}{10}\selectfont
   \ttfamily
   #1
  }%
}

\begin{document}

\title{The First Issue Matters: Linking Task-Level Characteristics to Long-Term Newcomer Retention in OSS}
% author here
\author{Yichen Hao}
\affiliation{
  \institution{
    School of Computer Science and Technology, 
    Soochow University}
  \city{Suzhou}
  \country{China}
}
\email{ychao02@stu.suda.edu.cn}

\author{Weiwei Xu}
\authornote{Corresponding Author}
\affiliation{
  \institution{
    School of Computer Science, Peking University}
  \city{Beijing}
  \country{China}
}
\email{xuww@stu.pku.edu.cn}

\author{Kai Gao}
\affiliation{
  \institution{
    School of Computer \& Communication Engineering, University of Science and Technology Beijing}
  \city{Beijing}
  \country{China}
}
\email{kai.gao@ustb.edu.cn}
\author{Xiaofang Zhang}
\authornotemark[1]
\affiliation{
  \institution{
    School of Computer Science and Technology, Soochow University}
  \city{Suzhou}
  \country{China}
}
\email{xfzhang@suda.edu.cn}
\begin{abstract}
    Sustaining newcomer participation is critical for the long-term health of open-source communities. Although prior research has explored various task recommendation approaches to help newcomers resolve their first-issue, these methods overlook how characteristics of first-issues may influence newcomers' long-term retention, limiting our understanding of whether initial success leads to sustained participation and hindering effective onboarding design. 
    In this paper, we conduct a large-scale empirical study to examine how first-issue characteristics affect newcomer retention. We combine predictive analysis, interpretability techniques, and causal inference to estimate the causal effects of issue characteristics on retention outcomes.
    The prediction task supports the interpretation and shows that interaction-related characteristics exhibit stronger associations with retention than intrinsic issue attributes. The causal analysis further reveals that issues reported by moderately experienced contributors, accompanied by moderate discussion intensity and participation from project members, and neutral or slightly negative comment sentiment, have higher retention potential.
    These findings provide actionable insights for OSS maintainers on designing issue management practices that better support long-term newcomer retention.

\end{abstract}

\maketitle
\section{Introduction}
In recent decades, open source software (OSS) projects have become a foundation of modern software development~\cite{li_SystematicLiterature_2024,hauge_AdoptionOpen_2010,nagle_LearningContributing_2018}, underpinning critical digital services and industrial ecosystems worldwide, from operating systems and frameworks to everyday and industrial applications~\cite{blind_EstimatingGDP_2024,feller_UnderstandingOpen_2002,fitzgerald_TransformationOpen_2006,_StateOpen_}. Their success is largely attributed to an open and collaborative development model~\cite{raymond_CathedralBazaar_1999,mockus_TwoCase_2002}, which enables developers worldwide to build, maintain and evolve complex software at an unprecedented scale~\cite{napoleao_OpenSource_2020,ghosh_UnderstandingFree_2005}.

Considering the importance of OSS, maintaining a healthy and active contributor community is essential for OSS projects~\cite{mohit_CommunityEngagement_2026}. In particular, the sustainability of OSS projects critically depends on attracting and retaining new contributors, as their open and voluntary nature necessitates a continuous influx of new contributors~\cite{zhou_WhatMake_2012,steinmacher_SystematicLiterature_2015}. Previous studies highlight that in most OSS projects, a small number of core developers shoulder the majority of development and maintenance effort, making these projects vulnerable to turnover~\cite{avelino_AbandonmentSurvival_2019,crowston_CorePeriphery_2006,crowston_SocialStructure_2005}. In this circumstance, keeping the influx of newcomers~\footnote{A newcomer, in the context of a particular OSS project (``repository'' on GitHub), refers to a developer who has not made any contributions to this repository.} greatly affects the long-term survival of OSS projects~\cite{xiao_HowEarly_2023,turzo_FirstPatch_2025}. 

However, from the newcomer's perspective, onboarding an OSS project and keeping the contribution can be quite difficult. Newcomers usually encounter substantial barriers when attempting to join an OSS project~\cite{steinmacher_AttractingOnboarding_2014}, including social obstacles, unfamiliar codebases and delayed or absent feedback from the community~\cite{steinmacher_SocialBarriers_2015,steinmacher_SystematicLiterature_2015}. In order to bridge the gap in \textit{making the first contribution}, GitHub provides several guidelines to build a healthy, newcomer-friendly community~\cite{_BuildingCommunities_}, along with the label ``\textbf{good first issue}'' (GFI) to mark issues that are suitable for newcomers~\cite{tan_FirstLook_2020}. Previous research also proposed various mechanisms of \textit{task recommendation} to help newcomers find the first-issue they can resolve, and further promotes their successful first contribution~\cite{xiao_RecommendingGood_2022,xiao_PersonalizedFirst_2023}. These methods make it easier for newcomers to find their entry point and make the first contribution.

% While existing task recommendation techniques and GFI labels successfully lower the initial entry barrier~\cite{steinmacher_SystematicLiterature_2015}, they primarily focus on helping newcomers complete their initial contribution. 
However, successfully making a first commit does not necessarily translate into sustained participation~\cite{steinmacher_AttractingOnboarding_2014,zhou_WhatMake_2012}; in fact, newcomers frequently abandon projects immediately after their first patch~\cite{zhou_WhoWill_2015,turzo_FirstPatch_2025,miller_WhyPeople_2019}. This highlights a critical oversight in current practices: existing mechanisms aim at easing onboarding, but they are not designed to optimize for long-term retention. A genuinely "good" first issue should serve not merely as an easy entry point, but as a catalyst for continuous engagement.

To better understand this limitation, it is important to examine how newcomers complete their first contribution. For many developers, their first contribution is a two-step process: selecting the entry task and actually resolving it~\cite{steinmacher_AttractingOnboarding_2014}. While previous studies have extensively shown that the social and technical interactions during the resolution process significantly affect retentions~\cite{zhou_WhoWill_2015,yue_GoodStart_2023,assavakamhaenghan_DoesFirst_2023}, the intrinsic influence of the entry task itself remains largely unexplored. We currently lack a systematic understanding of what characteristics make a first-issue capable of retaining its solver. This gap prevents maintainers from identifying or crafting issues that actively promote a sustainable influx of long-term contributors.

To bridge this gap, we design a large-scale empirical study to investigate the relationship between first-issue characteristics and newcomer retention, guided by the following research questions:
\begin{enumerate}
\item RQ1 (Prediction): To what extent can we predict a newcomer's retention outcome based solely on \textit{the observable context of the first issue}?

\item RQ2 (Interpretation): In the context of a first-issue, which intrinsic and interactive factors are most strongly associated with its retention potential?

\item RQ3 (Causal Effects): Beyond correlation, do these identified factors have a causal effect on newcomer retention after controlling for observed covariates?

\end{enumerate}

Specifically, we collect first-issue data from 1000 GitHub repositories along with the corresponding newcomer data, building a dataset with 45,865 samples of first-issue features, including issue description, issue comments \& events, issue reporter (i.e., the author of the issue) and details of the corresponding repository. \textbf{Stage~1.} We train an XGBoost classifier~\cite{chen_XGBoostScalable_2016} on our dataset to predict whether an issue can retain a newcomer after resolution. Based on the trained XGBoost model, we apply the SHAP explanation~\cite{lundberg_UnifiedApproach_2017,molnar_InterpretableMachine_2020} analysis on the model, and find out what features are most related to the ground truth label. This stage provides an initial understanding of our research target at the level of relevance. \textbf{Stage~2.} We apply causal inference according to the important factors in stage~1, and estimate their causal effects with further robustness examinations. 
% Note that, as a matter of fact, the ground truth label (retention outcome) in our dataset is determined by both issue factors and \textit{newcomer factors}, while the latter is unobservable in stage~1. So we supplement for newcomer features, including their historical activeness, contribution experience and account age, and control these factors as confounding factors. 
We employ Generalized Propensity Score (GPS) matching to adjust for observed covariates and estimate the dose-response function (DRF) to represent how a change in one factor has a causal effect on the outcome~\cite{hirano_PropensityScore_2004}.

Our research yields several important insights on what factors in the context of first-issues mostly affect the newcomer retention, which can indicate how to practically improve retention in GitHub projects. In SHAP explanation, we find that interaction-related factors in first-issues have the most relationship with retention outcome, which aligns with previous work~\cite{zhou_WhoWill_2015,tan_ItEnough_2023} and we further validate them with causality estimation. In the causal inference analysis, we obtain some notable findings: 1) \textbf{Interaction as early support}. Moderately intensive pre-implementation discussion boosts retention by reducing requirement uncertainty before coding begins, especially when project members participate in the discussion; 2) \textbf{Technicality over emotional positivity}. Sentiment positivity has a negative effect on retention, suggesting that newcomers benefit more from clear, technically-focused feedback than from overly emotional supportive but less informative responses. 3) \textbf{The expertise trade-off}. Reporter's experience exhibits a non-linear relationship with retention, where issues raised by highly experienced contributors can become less accessible to newcomers due to implicit assumptions and knowledge gaps.
In summary, this paper makes the following contributions.

% \vspace{-1mm}
\begin{itemize}
    \item A large-scale empirical study characterizing the retention patterns across diverse GitHub repositories with a machine learning mechanism to predict the retention potential of an issue;
    \item Key insights into issue-level factors influencing newcomer retention supported by causal analysis under observational assumptions, along with actionable recommendations for project maintainers.
    \item A comprehensive curated dataset comprising over 45,000 issue samples, integrating issue characteristics, newcomer attributes, and retention outcomes, which enables systematic empirical analysis of newcomer retention in OSS.
\end{itemize}

The remainder of this paper is organized as follows.
Section \ref{sec:related-work} overviews the previous research.
Section \ref{sec:rq} specifies the research questions.
Section \ref{sec:method} presents the methodology of the experiments.
Section \ref{sec:results} displays the results and analyzes the corresponding inspirations.
Section \ref{sec:threats} discusses the core enlightenment and threats to validity, and Section \ref{sec:conclusion} concludes this paper.

\section{Related work}
\label{sec:related-work}
\subsection{Factors of newcomer onboarding}
Extensive research~\cite{steinmacher_OvercomingOpen_2016,zhou_WhoWill_2015,steinmacher_SystematicLiterature_2015} has examined newcomer behaviors to uncover the factors that influence whether and how they successfully integrate into OSS communities.

Rehman et al.~\cite{rehman_NewcomerCandidate_2020} find that only about 30\% of those who attempt to onboard an OSS project finally succeed in making the first contribution, with primary barriers being identifying appropriate entry points and acquiring project-specific knowledge. Beyond technical barriers, Steinmacher et al.~\cite{steinmacher_SocialBarriers_2015} reveal that social communication difficulties often serve as the primary obstacles to newcomers’ initial participation. 

Focusing on contribution types, Subramanian et al.~\cite{subramanian_AnalyzingFirst_2022} find that newcomers tend to choose easy and newcomer-friendly tasks such as semantic bugs, small-scale patches, or documentation. However, simply assigning easy issues is insufficient to foster long-term engagement. Tan et al.~\cite{tan_ItEnough_2023} find that more than 60\% newcomers with future contributions are guided by experts during their first issue. Similarly, Turzo et al.~\cite{turzo_FirstPatch_2025} argue that while newcomer-oriented task recommendations help lower the initial entry barrier, they do not necessarily facilitate deeper understanding or sustained participation in OSS projects.

These studies collectively suggest that while newcomers’ initial task preferences are well documented, the mechanisms that transform early participation into sustained contribution remain unclear, motivating our investigation.

\subsection{Factors of newcomer retention}
Prior research on newcomer retention often focuses on long-term contributors to understand what enables newly joined developers to remain active in OSS projects~\cite{casalnuovo_DeveloperOnboarding_2015}. 
Studies show that timely feedback and positive interactions during a newcomer’s first contribution can significantly increase the retention rate~\cite{zhou_WhoWill_2015}, while project context also play a crucial role~\cite{bao_LargeScale_2021}. In addition, newcomer sentiment can affect retention and turnover status~\cite{tulili_ExploringTurnover_2025}. 

Empirical evidence suggests that the first contribution often determines future engagement. 
% Hence, identifying suitable first issues not only facilitates initial participation but also affects the sustained engagement of newcomers and the overall influx of active developers~\cite{xiao_PersonalizedFirst_2023}. Conversely, an unsuccessful first attempt may substantially reduce newcomers’ motivation to stay involved. Therefore, the selection of first tasks is not only a critical entry point for attracting and onboarding newcomers but also a potential key factor in promoting their long-term retention.
Therefore, identifying suitable first issues is not only a critical entry point for onboarding newcomers but also a key factor in promoting their sustained retention and the overall influx of active developers~\cite{xiao_PersonalizedFirst_2023}. 

\subsection{Predicting newcomer retention}
Zhou et al.~\cite{zhou_WhoWill_2015} developed predictive models based on newcomers’ early behaviors during their first contributions, demonstrating that such behavioral signals can effectively indicate future participation. Bao et al.~\cite{bao_LargeScale_2021} further incorporated features of both developers and OSS projects to predict whether newcomers would become long-term contributors. These approaches have revealed comprehensive factors that affect the retention of newcomers. 
However, most existing approaches focus primarily on individual-level features (e.g., newcomer's prior experience, activity patterns), while task-related characteristics (e.g., the nature of the first issue) are often overlooked. 

A parallel body of work examines task recommendation, including bug triaging and issue assignment~\cite{anvik_WhoShould_2006, canfora_HowSoftware_2005}. On platforms such as GitHub, maintainers are encouraged to label newcomer-friendly tasks using the \textit{good first issue (GFI)} tag~\cite{_EncouragingHelpful_}. Subsequent studies have leveraged GFI to recommend suitable entry tasks for newcomers~\cite{stanik_SimpleNLPBased_2018, tan_FirstLook_2020, xiao_RecommendingGood_2022, xiao_PersonalizedFirst_2023}. While GFI provides a valuable platform-level signal for analyzing task-side factors, these methods are primarily designed to facilitate initial onboarding rather than to predict or improve long-term retention.

Motivated by this gap, our work investigates large-scale first-issue data to examine which task-side factors are associated with newcomer retention during the initial contribution.

\section{Research Questions}
\label{sec:rq}

The core objective of this study is to understand whether certain issues are inherently more likely to facilitate future participation by newcomers. We aim to answer the following research questions.

    \paragraph{\textbf{RQ1}} \textit{To what extent can we predict a newcomer's retention outcome based solely on \textit{the observable context of the first issue}?}

    In task recommendation scenarios, previous work~\cite{xiao_RecommendingGood_2022,xiao_PersonalizedFirst_2023} aims at predicting whether an issue is suitable for newcomers based on its characteristics, yet it is not enough for retaining long-term contributors. The goal of this \textbf{RQ} is to verify that in the same real-world settings, the ability of classifier models to recognize first-issues that have the potential to retain newcomers.

    \paragraph{\textbf{RQ2}} \textit{In the context of a first-issue, which intrinsic and interactive factors are most strongly associated with its retention potential?}

    This \textbf{RQ} intends to interpret the classifier model in RQ1. While the model performance itself is informative, the prediction task is a means rather than an end, and it is worth exploring how the classification model makes the decision. Specifically, this \textbf{RQ} examines the characteristics of the first-issue that contribute the most to the model’s predictions, thereby revealing the factors most strongly associated with newcomer retention.

    \paragraph{\textbf{RQ3}} \textit{Beyond correlation, do these identified factors have a causal effect on newcomer retention after controlling for observed covariates?}

    The interpretation result shows the key characteristics for model prediction at the level of association, but it does not necessarily imply that changing these factors would causally influence newcomer retention~\cite{rudin_StopExplaining_2019}. Consequently, to obtain more rigorous insights, this \textbf{RQ} moves beyond correlation and aims to investigate the causal impact of these factors after adjusting for observed covariates.

\section{Methodology}
\label{sec:method}
In this section, we present our methodological framework adopted to investigate the \textbf{retention potential} of first issues in GitHub projects. 

\subsection{Experiment Implementation}
\label{method-def}

Our work intends to understand what first-issues have the potential to retain newcomers, and what characteristics of first-issues causally influence the retention outcome. Ideally, this would require a direct measurement of an issue's ``retention potential'', that is, whether an issue itself contains characteristics that encourage newcomers to continue contributing after resolving it. However, such a label is fundamentally unobservable in practice.

To implement this, we use a binary behavioral proxy: \textbf{whether a newcomer who resolved their first issue in a project proceeds to make further contributions}. While this proxy enables empirical study, it also introduces a critical challenge. An issue’s retention potential and a newcomer’s actual retention are not equivalent: newcomers’ subsequent contributions may be strongly influenced by factors beyond context of the issue itself, including motivation changes, external support, or personal circumstances. To ensure the validity of our conclusions, we first decompose the influence of newcomer factors into two distinct categories and address them accordingly.

    \textbf{Observable Systematic Factors}. Certain newcomer features, such as newcomer's overall OSS contribution experience and historical activity levels, can significantly impact retention probability in systematic level. These features cannot be adopted in \textit{practical prediction task} (where the newcomer is unknown ex-ante), but in the \textit{theoretical causal inference analysis} (where we seek to understand the true influence of issues), we can control these factors as \textbf{confounders} to estimate causal effect of issue's factors.
    
    \textbf{Unobservable Stochastic Factors}. Still, factors like newcomer's personal schedule or circumstances are unable to observe from GitHub data source, but these factors act as \textbf{random, non-systematic noise}. Given the large scale of our dataset, we assume these variations are distributed stochastically and do not fundamentally skew the directional causal relationship between issue features and retention potential. For potentially influential unmeasured confounders that may violate this assumption, we can assess the robustness of our estimates using an E-value sensitivity analysis~\cite{vanderweele_SensitivityAnalysis_2017}.

This decomposition helps mitigate bias in the empirical analysis. Driven by the \textbf{RQs}, we design our experiments as follows.

\vspace{-1mm}
\subsubsection{Stage 1: Correlation}
    \textbf{Exp1: Prediction}. This experiment addresses practical scenario and trains a classification model on the context of first issue. This stage validates whether issue characteristics alone hold sufficient predictive power, despite those unobservable factors in advance. \textbf{Exp2: Interpretation}. We perform interpretable analysis on the trained model to draw preliminary conclusions about the \textit{correlation} between issue features and retention potential.
\vspace{-1mm}
\subsubsection{Stage 2: Causality}
    \textbf{Exp3: Causal Inference}. Based on the findings in the interpretable analysis, we adopt causal inference methods to examine whether these correlations can be escalated to the level of causality. Specifically, we estimate causal effects by adjusting for a set of observed covariates capturing systematic factors defined above. While unobserved confounding may still exist, this approach helps reduce confounding bias and provides evidence on the causal effects of issue characteristics on retention potential.
\vspace{-1mm}
\subsection{Data Collection}
\label{subsec:datac}
To understand how intrinsic factors of an issue affect its retention potential in real-world OSS projects, we build our dataset based on data from GitHub repositories. The construction of our dataset contains two phases: 1) collecting raw data of first-issues from GitHub and 2) collecting \& calculating detailed feature fields for raw data based on feature engineering. In this section, we introduce the process of the first phase, and discuss the second phase along with feature extraction in next section.

Considering that GHTorrent, a popular dataset of GitHub archive, has not been maintained for several years, we collect metadata through official GitHub REST API~\cite{_GitHubREST_} and GraphQL API~\cite{_GitHubGraphQL_}. The specific steps are as follows.

    \subsubsection{Repositories} \label{subsec:repoc} We filter a series of GitHub repositories as the initial data scope, following these criteria: \textit{has no less than 200 stars, 5 forks and 10 open issues and is updated since 2024-01-01}~\footnote{The data are retrieved on Feb 2025.}. We employ the search function in GitHub GraphQL API to get metadata of repositories which fulfill the criteria, and record the repository-id (e.g. microsoft/vscode) along with the count of stars \& forks of the repositories in the database. The search API of GitHub returns a maximum of 1,000 results at one time, and we keep just the first 1,000 returned repositories since we observe in the following steps that the first-issue samples in these repositories are already enough for our empirical study.
\vspace{-1mm}
    \subsubsection{Issues \& Resolvers} For each repository, we use GraphQL API to collect all the issues closed before 2024-01-01, and record the issue number and basic information including issue content, created time and closed time. 
    % However, not all issues are intended for software development tasks, and we need to filter the issues which have a specific resolver. To implement this, we check whether an issue is closed by a Pull Request (PR) or a Commit. On GitHub, an issue can be referenced by a PR or Commit via text patterns such as ``close(s)/fix(es)/resolve(s) \#issue\_num'' in the PR title or description / Commit message, and this reference can be extracted from the \Code{timelineItems $\rightarrow$ ClosedEvents} node in GitHub GraphQL API. We discard those issues that do not have any \textit{CloserEvent} in the query result, while for the remaining samples, the resolver is just the author of the corresponding PR or Commit, and we also record the data of the PR or Commit. 
    However, not all issues are intended for development tasks. To obtain a specific resolver for each issue, we only retain the issues that are closed by a Pull Request (PR) or a Commit, and the author of the PR or Commit is the corresponding resolver.
    Some repositories do not retain any issue samples after this filtering step and are therefore excluded, leaving 777 repositories.

    \subsubsection{First issues \& Newcomers} Among all the issues retrieved along with the resolver of each issue, we further recognize the \textbf{first issue} samples and the corresponding \textbf{newcomers}. 
    Our identification is conducted at the repository level and follows a two-step procedure.

    \textbf{First}, for each repository, issues are ordered by closing time, and an issue is labeled as a \textit{first issue} if its resolver appears for the first time when resolving that issue.

    \textbf{Second}, to ensure that such resolvers have no prior involvement, we exclude those with any earlier contributions (e.g., commits or pull requests) to the repository before the issue’s closing time.
    % Specifically, we query the GitHub GraphQL API to retrieve the number of commits and pull requests made by the resolver to the repository before resolving the issue. 
    % Only issues whose resolvers have no prior contributions (commits or PRs) to the repository are retained.
    
    This process yields a set of \textbf{first issues}, each associated with a \textbf{newcomer}, defined as a developer whose first contribution to the repository is resolving that issue.

    \subsubsection{Newcomer Retention Outcome (label)} As defined in Section~\ref{method-def}, we use the observable \textit{retention outcome} as the proxy of issue's retention potential. For newcomers in a repository, we search for the number of their future contributions (including PRs and Commits) in this repository. Based on this, we can classify the newcomers as \textit{retaining} (label = 1) and \textit{not retaining} (label = 0) according to whether they make more than \textit{k} future contributions after resolving the first issue. The \textit{k} here is a threshold parameter that can be controlled in experiments to satisfy different scenarios.

Through these steps, we construct the raw data of first issues and newcomers of our dataset, containing 777 repositories and 45,865 first issues in total.
\vspace{-1mm}
% \subsection{Feature Extraction}
% To precisely describe the factors of a first issue that might affect its retention potential, we extract features from these dimensions.
% Our feature design is largely informed by prior empirical studies on issue resolution, developer participation, and repository backgrounds in OSS projects~\cite{zhou_WhoWill_2015,xiao_RecommendingGood_2022,xiao_HowEarly_2023}, which are used for the prediction task. We also extract a set of features of newcomers to control their confounding effects in the causal inference stage.

% According to the real-world task recommendation settings, feature variables used in our experiments should be observable in the issue \textbf{before the participation of the corresponding newcomer}. To implement this, we define a \textbf{cutoff timestamp} and ensure to use the data observed at this timestamp. Considering that in most cases, developers resolve an issue by opening a pull request in the GitHub repository~\cite{gousios_ExploratoryStudy_2014,tsay_InfluenceSocial_2014}, we use the \textit{creation time of this pull request} as the cutoff timestamp for this issue; if an issue is not referenced by a PR, we use the \textit{closing time of this issue} instead. 
\subsection{Feature Extraction}
To characterize the factors of a first issue that may relate to newcomer retention, we extract features from multiple dimensions.
Our feature design is largely informed by prior empirical studies on issue resolution, developer participation, and repository backgrounds in OSS projects~\cite{zhou_WhoWill_2015,xiao_RecommendingGood_2022,xiao_HowEarly_2023}. These features are used in the prediction task. We also extract a set of newcomer-level features to control for potential confounding effects in the causal inference stage.

For practical prediction settings, the feature variables used in our experiments should be observable before the resolution of the focal issue is completed. To ensure this, we define a \textbf{cutoff timestamp} and only use data available up to that time. Because developers typically resolve GitHub issues by submitting a pull request~\cite{gousios_ExploratoryStudy_2014,tsay_InfluenceSocial_2014}, we use the \textit{creation time of the resolving pull request} as the cutoff timestamp whenever such a pull request is linked to the issue. For resolved issues without a referenced pull request, we use the \textit{issue closing time} as the cutoff timestamp instead.
% Specifically, the cutoff timestamp can be obtained from the raw data for each sample. In collecting the remaining data of feature variables, most interfaces provided by GitHub REST API and GraphQL API support \textit{time-based filtering} or \textit{attaching timestamps to retrieved events}, so we leverage this property to collect the feature data observed at the cutoff timestamp. 
% By implementing this, we retrieve the feature data to build a dataset that restores the issue samples to the cutoff timestamp, thereby meeting the real-world task recommendation scenarios.
\vspace{-1mm}
\subsubsection{Issue contents}This dimension captures the intrinsic properties of an issue as reflected in its textual description.
Previous studies~\cite{xiao_RecommendingGood_2022,panichella_WontWe_2021,ehsani_WhatCharacteristics_2025} have found that, the text length of issues (\Code{len\_title}, \Code{len\_body}), the readability of issue description (measured by \textit{readability index} as 
% \Code{flesch\_reading\_ease}, \Code{automated\allowbreak\_readability\allowbreak\_index}, \Code{difficult\_words} and \Code{linsear\_write\_formula}) 
flesch\_reading\_ease, automated\_readability\_index, difficult\_words and linsear\_write\_formula)
and the richness of issue description (\Code{len\_code}, \Code{\#codes} as the number of code snippets, \Code{\#imgs} as the number of images) are the basic metrics to characterize the content of an issue. In resolving the issue, developers can manually add \textit{labels} to classify the issue into specific categories, so we refer to previous work~\cite{xiao_RecommendingGood_2022} and divide common labels into 12 categories (i.e., \textit{Bug, Documentation, Test, Build, Enhancement, Coding, New Feature, GFI-Signaling, Medium Difficulty, Difficult/Important, Triaged and Untriaged}) using a rule-based method.

% Note that we also compute metrics such as \textit{high-frequency words} by TF-IDF~\cite{sparckjones_STATISTICALINTERPRETATION_1972} and \textit{text embedding} by adopting \Code{sentence-transformers} framework~\cite{reimers_SentenceEmbeddings_2019} and \Code{all-MiniLM-L6-v2} model to capture semantic information of textual content of issues, but this does not work well under our prediction task, so these metrics are not introduced in the main experiment of the prediction task.
\vspace{-1mm}
\subsubsection{Interactions within issue}
This dimension characterizes the social and collaborative dynamics within an issue discussion.
We derive interaction-related features that describe the intensity, timeliness, and structure of communication between participants.

Specifically, we extract and compute these features for characterizing issue comments: the total number of comments, the number of comments sent by the contributors of the repository, the number of participants who are engaged in the comments, and the sentiment metrics of the comment text using SentiStrength-SE~\cite{islam_SentiStrengthSEExploiting_2018}.
For the events in issues, we simply compute the total number of events, and the number of events triggered by the contributors of the repository.
\vspace{-4mm}
\subsubsection{Background--Issue reporter}
In previous research~\cite{xiao_RecommendingGood_2022,xiao_PersonalizedFirst_2023}, factors of \textit{reporters} (i.e., the author of the issue) have been shown to be important in predicting whether an issue is suitable for newcomers. So we also adopt this dimension and extract features about the reporters.
To measure the overall development experience of reporters, we record \textit{the number of commits / PRs / issues} contributed by the reporter. We also record the contribution statistics of the reporter within the corresponding repository of the issue, to reveal the reporter's contribution level in this repository.
In particular, we care about the actual impact brought by reporters, so we compute the following 2 metrics: the number of issues reported by the reporter in the corresponding repository that \textit{1) are resolved by newcomers}, and \textit{2) are resolved by newcomers who retained}.
\vspace{-1mm}
\subsubsection{Background--Repository}
Finally, considering that issues in different repositories might have different intrinsic properties, we include repository-level features to control for structural differences across projects.

We characterize the repositories in two aspects: \textit{the basic information of the repository} (the number of \textit{stars}, the number of historical contribution, and repository age), and \textit{the historical development dynamics in the repository} (the average time spent on resolving issues, the number / percent of issues resolved by newcomers, the number / percent of issues that retained newcomers, the average number of PR reviews, and sentiment metrics of PR review comments). The former describes the basic features of a repository, while the latter characterize the development patterns in this repository.

\subsubsection{Background--Newcomer (for causal control)}
In observational OSS data, retention outcomes are jointly influenced by both issue characteristics and individual newcomer attributes.
Failing to control for such individual differences may introduce confounding bias when estimating the causal effects of issue properties, so it is necessary to extract \textbf{the features of newcomers}. These features are not used in the prediction task in prevention of data leakage.

According to previous work~\cite{xiao_HowEarly_2023,zhou_WhoWill_2015}, we use the following metrics to characterize a newcomer: GitHub account age, activeness (i.e., the number of contributions within 3 / 6 months prior to issue resolution), overall experience (i.e., the number of issue-reporting / commits / PRs / total contributions) and the number of repositories owned by the newcomer.

\vspace{-1mm}
\subsection{Predictive Modeling: XGBoost \& SHAP}
To address the prediction task in Stage~1, we formulate newcomer retention prediction as a binary classification problem. For each first issue, the model predicts whether its resolver (newcomer) will be retained in the repository, using only \textit{ex-ante} issue context that is observable at the time before the issue is resolved.

\subsubsection{Model choice}
We adopt XGBoost as the predictive model due to its strong performance on tabular data and ability to capture non-linear feature interactions~\cite{chen_XGBoostScalable_2016}. XGBoost has been widely used in empirical software engineering studies for predictive tasks, making it a suitable choice for our setting.

\subsubsection{Training and evaluation}
\label{method-train}
We train the model on the first-issue dataset and evaluate it using standard classification metrics. By varying the threshold \textit{k}, we examine performance under different retention levels: positive samples (label = 1) correspond to newcomers retained for at least \textit{k} subsequent contributions, while negative samples are limited to those who made no further commits or pull requests after resolving their first issue. Table~\ref{tab:performance_k} reports the selected \textit{k} values and the corresponding sample sizes.

As \textit{k} increases, the class distribution becomes more imbalanced. To address this, we use 10-fold cross-validation, balancing only the training folds by resampling while keeping each test fold unchanged to preserve the original distribution in the real-world. We report the average performance over the 10 folds.

%It is obvious that the number of positive and negative samples becomes imbalanced as \textit{k} increases, which is inappropriate for training a binary classifier model. As a result, we propose our approach for dataset splitting. 

%We use 10-fold cross-validation to evaluate the predictive performance in a robust manner. 
% In particular, the dataset is randomly partitioned into 10 mutually exclusive folds with equal sizes. In each iteration, one fold is held as the test set, while the other nine folds are used for training the model. 
%To address the imbalance, we resample the training folds to build the training set and ensure that the number of positive and negative samples are the same in the training set for each fold iteration. The test set is left unchanged in each iteration to reflect the origin distribution in the real-world. The classifier is trained on the balanced training set and evaluated on the origin test fold for 10 iterations, and reports the average performance to reduce variance caused by data splitting.
\vspace{-1mm}
\subsubsection{Interpretable analysis}
To understand how issue-level features contribute to the model’s predictions, we apply SHAP to the trained XGBoost model. 
SHAP is a model-agnostic interpretability method grounded in Shapley values from cooperative game theory, providing additive feature attribution scores for individual predictions~\cite{lundberg_UnifiedApproach_2017}. 
For a given instance, SHAP models the prediction task as a cooperative game in which each feature represents a “player,” and the model output corresponds to the payoff. The \Code{SHAP value} of a feature is defined as its average marginal contribution to the prediction across all possible feature subsets, ensuring a fair attribution of the prediction among features. This formulation satisfies desirable theoretical properties such as \textbf{local accuracy}, \textbf{consistency}, and \textbf{missingness}, which collectively guarantee that the sum of feature attributions equals the model prediction and that more influential features receive larger attributions~\cite{lundberg_UnifiedApproach_2017}. In practice, SHAP assigns a \Code{SHAP value} to each feature for a given sample, indicating its contribution to the prediction. To obtain global feature importance, SHAP aggregates these contributions across all samples by computing the mean absolute SHAP value of each feature.

% SHAP naturally supports \textbf{local interpretability} by explaining individual predictions through instance-specific feature attributions. To derive \textbf{global insights}, SHAP values can be aggregated across multiple instances, enabling the analysis of overall feature importance and the identification of systematic patterns in model behavior. In practice, summary statistics (e.g., mean absolute SHAP values) and visualization techniques are commonly employed to characterize the global influence of features while preserving the consistency between local and global explanations.

In our study, SHAP is used to reveal \textit{associative patterns} between issue characteristics and retention outcomes rather than causal effects. 
The insights obtained from SHAP serve as preliminary evidence and motivate the causal analysis conducted in Stage~2.

\vspace{-1mm}
\subsection{Causal Inference}
\subsubsection{Motivation and causal question}
The interpretable analysis in Stage~1 is able to reveal the feature variables that are strongly associated with newcomer retention. However, such associations do not necessarily imply causality, as they may arise from confounding factors that jointly influence both issue features themselves and retention outcomes. Therefore, in Stage~2, we explicitly investigate whether the identified issue characteristics have causal effects on retention.

Formally, our causal question is: \emph{given two issues that differ in one specific feature variable while being comparable in all other relevant aspects, how does this difference causally affect the probability that a newcomer is retained by one issue?}

\vspace{-2mm}
\subsubsection{Treatment, outcome, and confounders}
\label{causal-def}
In causal analysis, three fundamental concepts are defined as follows: the \emph{treatment} represents the variable whose causal effect we aim to estimate; the \emph{outcome} is the response or effect potentially influenced by the treatment; and \emph{confounders} are variables that affect both the treatment and the outcome, potentially biasing causal estimates if not properly controlled.

In this study, the treatment variable corresponds to continuous-valued feature variables identified as important in Stage~1. The outcome variable is the binary retention indicator of the newcomer who first engages with the issue. We further incorporate \emph{newcomer-related features} which capture systematic differences among newcomers, and treat them as confounders since they may simultaneously influence both issue selection and retention outcomes.

\subsubsection{Generalized propensity score for continuous treatments}
Unlike classical propensity score methods designed for binary treatments~\cite{rosenbaum_CentralRole_1983}, issue characteristics in our setting are inherently continuous. To accommodate this property, we adopt the Generalized Propensity Score (GPS) framework~\cite{hirano_PropensityScore_2004}. 
The GPS summarizes the conditional distribution of a continuous treatment given confounders, and \textbf{enables unbiased estimation} of causal effects by balancing confounding covariates across different treatment levels.

Specifically, let $T_i$ denote the continuous treatment variable for the $i$-th instance and $X_i$ be the vector of observed confounders. We first apply a logarithmic transformation, $\tilde{T}_i = \log(T_i + 1)$, to mitigate skewness. Following prior work on continuous-treatment causal inference~\cite{hirano_PropensityScore_2004}, we assume the transformed treatment follows a conditional normal distribution given the covariates:
\begin{equation}
    \tilde{T}_i \mid X_i \sim \mathcal{N}(\beta^T X_i, \sigma^2).
\end{equation}
The parameters $\beta$ and $\sigma^2$ are estimated via \textit{ordinary least squares} (OLS)~\cite{garthwaite_StatisticalInference_2002}, which corresponds in principle to a \textit{linear regression model}. The GPS, denoted as $R_i$, is then calculated as the value of the probability density function at the observed treatment level:
\begin{equation}
    \hat{R}_i = \frac{1}{\sqrt{2\pi\hat{\sigma}^2}} \exp \left( -\frac{(\tilde{T}_i - \hat{\beta}^T X_i)^2}{2\hat{\sigma}^2} \right).
\end{equation}
To ensure the validity of causal inference, we assess the covariate balance using stabilized inverse probability weighting derived from $\hat{R}_i$. We verify that the absolute correlations between the confounders and the treatment are negligible after adjustment ($|corr|<0.1$). These diagnostics suggest that weighting substantially reduces observed confounding.
Based on the verified sample, we subsequently estimate the dose--response function to quantify how varying treatment levels affect the outcome.

\subsubsection{Dose--response function estimation}
% Based on the estimated GPS, we further estimate the dose--response function (DRF), which characterizes how the expected retention outcome varies as a function of the treatment level. The DRF provides a fine-grained causal interpretation, allowing us to quantify not only whether an issue characteristic causally affects retention, but also how changes in its magnitude influence retention probability.
To quantify how changes in issue characteristics influence retention, we estimate the dose--response function (DRF), $\mu(t) = E[Y(\tilde{t})]$, where $Y$ is the retention outcome. Following the GPS adjustment approach, we model the conditional expectation of the outcome as a quadratic function of the treatment and the estimated GPS:
\begin{equation}
    E[Y_i \mid \tilde{T}_i, \hat{R}_i] = \alpha_0 + \alpha_1 \tilde{T}_i + \alpha_2 \tilde{T}_i^2 + \alpha_3 \hat{R}_i + \alpha_4 \hat{R}_i^2.
\end{equation}
% We exclude interaction terms based on model fit diagnostics. 
% Our causal interpretation relies on the standard assumptions of weak unconfoundedness and overlap in continuous-treatment causal inference~\cite{hirano_EfficientEstimation_2003}. To support these assumptions, we include a comprehensive set of covariates identified in prior studies and restrict the estimation to the common support region of the treatment distribution.
Finally, the causal effect at a specific treatment level $t$ is obtained by averaging the predicted outcomes over the distribution of the GPS across the entire population:
\begin{equation}
    \hat{\mu}(t) = \frac{1}{N} \sum_{i=1}^{N} \left( \hat{\alpha}_0 + \hat{\alpha}_1 \tilde{t} + \hat{\alpha}_2 \tilde{t}^2 + \hat{\alpha}_3 \hat{R}_i(t) + \hat{\alpha}_4 \hat{R}_i(t)^2 \right),
\end{equation}
where $\tilde{t} = \log(t+1)$ and $\hat{R}_i(t)$ represents the counterfactual GPS for unit $i$ evaluated at treatment level $\tilde{t}$. This integration step removes the dependency on specific covariate values, isolating the marginal causal effect of the treatment.

\section{Results}
\label{sec:results}

This section displays the results of experiments, answers the research questions and discusses practical inspirations.

\subsection{RQ1: Performance of Prediction Task}
\subsubsection{Metrics}
We use \textbf{AUC} (Area under the ROC Curve), \textbf{Accuracy} and \textbf{Recall} to evaluate the predictive performance of classifier models.

The classifier model outputs a probability for our binary prediction task and then converts to the binary 0/1 label according to a classification threshold (usually default to 0.5). Metrics such as \texttt{accuracy, precision} and \texttt{recall} rely on the binary label, which can be influenced by the choice of the threshold. \texttt{AUC}, in contrast, evaluates the model’s overall discriminative ability by measuring the probability that \textit{a randomly chosen positive instance is ranked higher than a randomly chosen negative instance}, independent of any specific classification threshold. As a threshold-agnostic metric, AUC provides a more robust and comprehensive assessment of model performance under class imbalance, which is particularly important in our setting where the ratio between retained and non-retained newcomers varies with the retention threshold~$k$ proposed in Section~\ref{method-train}. \texttt{Accuracy} and \texttt{Recall} are also used under the classification threshold of 0.4 for reference.

\subsubsection{Performance under different retention threshold~$k$}

\begin{table}[t]
\centering
\footnotesize
\caption{Classification performance and class distribution under different retention thresholds $k$.}
\vspace{-2mm}
\label{tab:performance_k}
\begin{tabular}{cccccc}
\toprule
\multirow{2}{*}{$k$} & \multicolumn{3}{c}{Performance} & \multicolumn{2}{c}{Class Distribution} \\
\cmidrule(lr){2-4} \cmidrule(lr){5-6}
 & ACC & AUC & Recall & \# Positive & \# Negative \\
\midrule
1  & 0.596 & 0.650 & \textbf{0.761} & 21,845 & \multirow{6}{*}{24,020} \\
2  & 0.603 & 0.681 & 0.742 & 16,576 &  \\
4  & 0.637 & 0.726 & 0.721 & 9,977 &  \\
6  & 0.665 & 0.752 & 0.717 & 7,432  &  \\
8  & 0.690 & 0.766 & 0.698 & 6,121  &  \\
10 & \textbf{0.704} & \textbf{0.771} & 0.687 & 5,257  &  \\
\bottomrule
\end{tabular}
\end{table}

Table~\ref{tab:performance_k} shows the results of our prediction task under different retention threshold~$k$, which reveals the model capability of distinguishing retained / non-retained newcomers under different retention levels.

% In general, the prediction performance improves as $k$ increases, indicating that under higher $k$, the classification task becomes easier mainly due to the growing separability between retained newcomers and non-retained newcomers, which leads to improvements in ACC and AUC. However, the recall gradually decreases as $k$ increases. Since recall measures the model’s ability to identify positive cases (retained newcomers), this pattern suggests that a stricter retention threshold makes the positive class harder to capture, likely because retained newcomers become rarer under larger $k$.
Overall, the consistently increasing ACC and AUC across different $k$ indicates that the model possesses a promising capability in distinguishing first issues with different retention potentials, suggesting its potential usefulness in retention-oriented issue recommendation scenarios.

To ensure that the predictive signal is not specific to XGBoost, we compare it with baselines (linear regression, random forest~\cite{breiman_RandomForests_2001} and TabPFN~\cite{hollmann_TabPFNTransformer_2023}). Although the baselines achieve lower absolute performance, they exhibit similar performance trends as $k$ varies. This suggests that the main pattern underlying our subsequent interpretability analysis is not an artifact of a single model. 

\subsubsection{Role of the Prediction Task}
The target of the prediction task is \textbf{not to obtain an excellent classification accuracy}. We design this prediction task to simulate a realistic issue recommendation scenario, but as mentioned in Section \ref{method-def}, predicting the retention potential of first issues solely by ex-ante issue-related features is inherently \textit{challenging} in the real world settings. From this perspective, the moderate predictive performance observed in RQ1 reflects the intrinsic difficulty of the problem rather than deficiencies in modeling or feature engineering. 
% While incorporating richer behavioral or post-hoc signals may further improve prediction accuracy, such extensions fall outside the scope of this study.

More importantly, the prediction task in RQ1 serves as a means rather than an end. Instead of aiming to directly solve the retention prediction problem in a single step, we use this experiment to systematically identify which issue-related factors are most strongly associated with retention outcomes. These findings then motivate the interpretability analysis in RQ2 and enable a principled causal investigation in RQ3. 

\begin{rqsummary}
    \textbf{Summary for RQ1.}
    Predicting newcomer retention based solely on ex-ante, issue-related features is a challenging task, and the achieved performance remains moderate. Nevertheless, the results demonstrate that issue characteristics contain non-trivial signals about retention potential, and establish the empirical foundation for the subsequent interpretability analysis and causal inference.
    % , rather than serving as an end goal of predictive optimization.
    
\end{rqsummary}

\subsection{RQ2: Interpretability Analysis}
Based on the trained XGBoost classifier model in RQ1, we employ the \Code{TreeExplainer} in Python \Code{shap} package to explain the model, since \Code{TreeExplainer} computes SHAP values efficiently and exactly for tree-based models using the TreeSHAP algorithm~\cite{lundberg_LocalExplanations_2020}.
This interpretability analysis enables both instance-level and aggregated explanations by attributing each prediction to individual feature contributions, thereby revealing the relative importance, directional effects, and stability of different feature categories across folds.
Specifically, we used the trained XGBoost classifier model under the threshold $k=6$, where the model performance is reasonable and representative.

\subsubsection{Important features}
\begin{figure*}[t]
  \centering
  \includegraphics[width=0.7\linewidth]{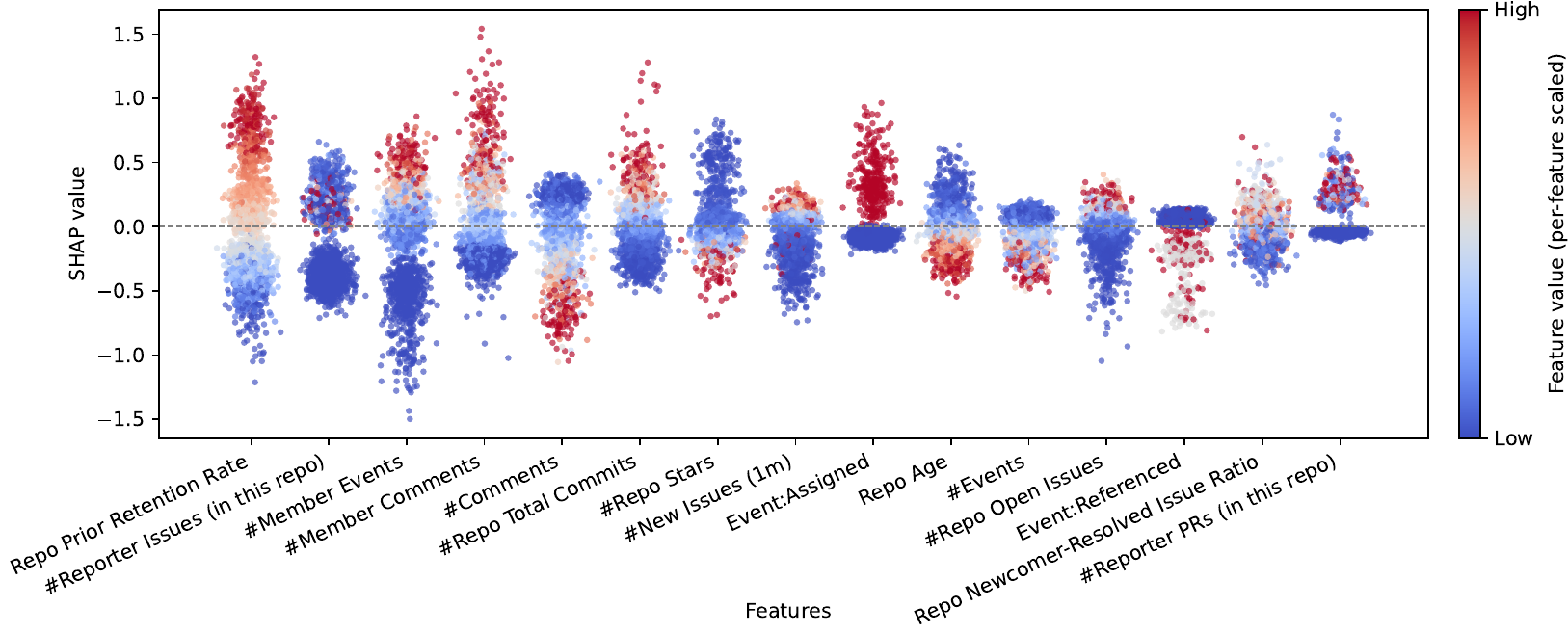}
  
  \vspace{-3mm} \caption{SHAP summary plot. Features are ordered from left to right by decreasing mean absolute SHAP value. Each point represents a sample and color indicates the feature value of this sample.}
  \vspace{-3mm}
  \label{fig:shap-summary}
  \Description{A wide rotated SHAP summary beeswarm plot for two-column layout.}
  
\end{figure*}
Figure \ref{fig:shap-summary} shows the summary of the interpretability analysis based on SHAP.
Overall, the SHAP results exhibit a clear and consistent pattern. Features related to \textbf{social interaction during issue resolution} and \textbf{project-level historical context} dominate the model’s explanations. In particular, interaction-oriented variables, such as the number of member comments and events within the issue, as well as the reporter’s prior activity, show the highest mean absolute SHAP values. 
% These features also demonstrate strong and stable directional effects: across all cross-validation folds, higher values consistently contribute positively to the predicted retention probability, with large effect sizes when comparing instances with positive and negative SHAP attributions.

In contrast, features describing the \textbf{intrinsic properties of the issue itself}, including textual length and readability proxies, contribute little to the model’s predictions. Their SHAP values remain close to zero, and the difference in feature distributions between positively and negatively attributed instances is small and unstable. This observation is robust across folds, suggesting that the limited contribution of issue content features is not an artifact of a particular training split, but rather a systematic pattern under the current prediction setting.

\subsubsection{Insights of the newcomer retention problem}
Taken together, these findings challenge the intuitive expectation that the quality or completeness of a first issue’s description is the primary driver of newcomer retention. Instead, the model consistently attributes retention potential to factors reflecting \textit{how newcomers are received and supported during the issue-solving process}, as well as to the broader ecosystem characteristics of the hosting repository. From this perspective, a first issue functions less as a static task description and \textbf{more as a trigger for early social interaction}. 
% The presence of timely responses, engagement from experienced members, and a historically supportive project environment appears to play a more prominent role in encouraging continued participation.

% Nevertheless, it is important to interpret these results with caution. SHAP-based explanations characterize associations learned by the predictive model rather than causal effects. In particular, the weak attribution of issue content features may be influenced by masking effects or nonlinear interactions with stronger social signals. Given that prior software engineering studies emphasize the importance of documentation quality and information completeness for newcomer onboarding~\cite{tan_FirstLook_2020,xiao_RecommendingGood_2022,turzo_FirstPatch_2025}, the apparent insignificance of issue content in our interpretability analysis warrants further scrutiny.
% Therefore, while RQ2 reveals that the model predominantly relies on social and contextual signals when estimating retention potential, it does not conclusively rule out the \textbf{causal relevance} of issue-intrinsic characteristics. 
To disentangle causality from correlation, we proceed to RQ3, where we apply causal inference techniques to explicitly evaluate the effects of selected issue characteristics on newcomer retention.

\begin{rqsummary}
\textbf{Summary for RQ2.}
    Our interpretability analysis shows that retention predictions are driven primarily by social interaction and project-level context, rather than by the intrinsic characteristics of the first issue. This finding suggests that early social engagement, rather than issue description quality, plays a central role in newcomer retention.
\end{rqsummary}

\subsection{RQ3: Causal Effects}

In RQ1 and RQ2, we formulate newcomer retention as a prediction task in a realistic issue recommendation setting, using only ex-ante, issue-related features. While this formulation aligns with the information constraints of practical recommender systems, it confounds issue characteristics with the competence and behavior of the newcomers who ultimately resolve the issues. Consequently, purely predictive evidence cannot exclude an alternative explanation: the observed patterns may primarily reflect self-selection of more capable newcomers, rather than effects attributable to the issues themselves.
\begin{figure}[t]

    \centering
    \includegraphics[width=\linewidth]{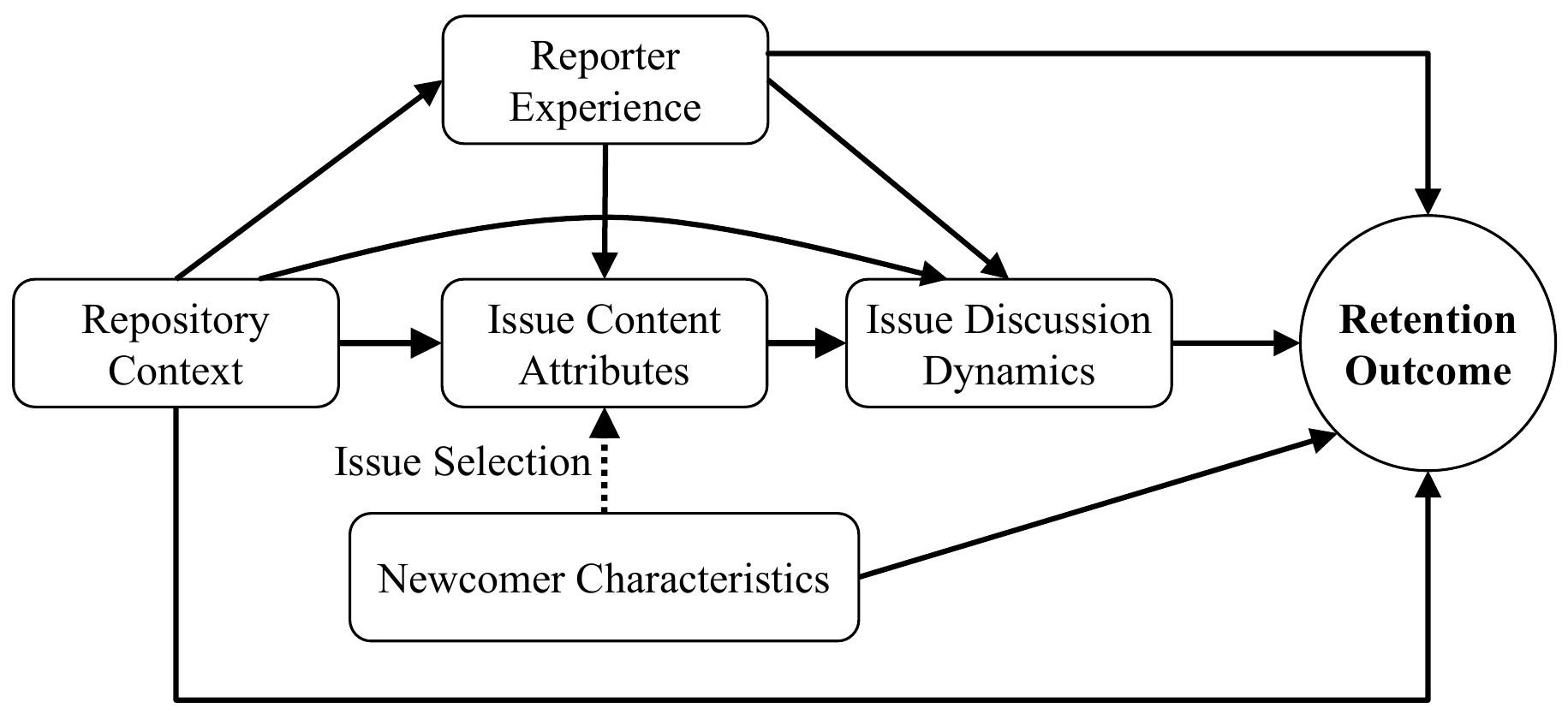}
    \vspace{-3mm}
    \caption{Causal DAG for modeling issue-related factors and newcomer retention. Treatments are instantiated from the graph, and covariates are selected via the backdoor criterion.}
    \vspace{-3mm}
    \label{fig:causal-dag}
\end{figure}

RQ3 addresses this limitation through a causal inference lens. 
% Building on the same feature data as RQ1 and RQ2, we additionally control for newcomer-specific factors as global confounders, along with specific covariates for each treatment, enabling estimation of the causal effects of issue characteristics on retention. 
Building on the same feature data as RQ1 and RQ2, we additionally obtain newcomer-specific factors for confounder adjustment.
Figure~\ref{fig:causal-dag} presents the \textbf{Causal Directed Acyclic Diagram} (DAG) of our study, capturing a temporally ordered data-generating process where all variables are measured prior to the newcomer's first contribution, thus avoiding reverse causality. Repository context, reporter experience, and issue content jointly influence discussion dynamics (treatment) and may also directly affect retention (outcome). For newcomer characteristics, while they do not directly affect the issue, newcomers with different backgrounds may choose different issues, thus introducing \textit{selection bias}.
Following this DAG, we select covariates based on the backdoor criterion: for each treatment, we adjust for upstream variables that affect both the treatment and outcome, while excluding post-treatment variables. Newcomer characteristics are additionally controlled to mitigate selection effects.
% 重复Importantly, all treatments considered in this section are observable prior to any code contribution or pull request submission, preserving the applicability of our findings to real-world issue triage and recommendation.

To examine whether our findings are sensitive to the choice of retention threshold $k$, we conduct each causal inference analysis on datasets constructed with $k=1,2,4$ and $6$, and report the results across all four datasets. This design reduces the risk that the observed effects are tied to one particular $k$, and thus provides stronger evidence for the robustness and credibility of our findings.

\subsubsection{Issue Discussion Dynamics}
The importance of early discussion dynamics in a newcomer’s first issue has been established in prior work~\cite{zhou_WhoWill_2015,assavakamhaenghan_DoesFirst_2023} and echoed in our RQ2 results; here, we further examine its causal effect.
% According to the definition in Section~\ref{causal-def}, causal effect estimation requires explicit control of confounding variables when analyzing the relationship between a treatment and the retention outcome. To obtain estimated causal effect between discussion dynamics (through comments) of an issue and its retention potential, we adjust for these groups of covariates.
% First, we control for characteristics of \textbf{issue description content}, because they can affect the discussion patterns in an issue and influence the retention potential of the issue as well.
% Second, we control for \textbf{repository-level characteristics} (i.e., maturity and historical activity), because repositories with different onboarding norms and process maturity can systematically affect retention via project complexity and support capacity.
% Third, we control for \textbf{reporter-level characteristics}, as issues authored by experienced developers in the corresponding repository may involve more in-depth discussions and may also indirectly improve retention through better issue triage and faster follow-up.
% Finally, we control for \textbf{newcomer-level characteristics}, because more capable or more active newcomers may select into issues with discussions that reflect a challenging problem, and are also more likely to remain in the project regardless of the discussions in the issue.
% Controlling for these covariates allows us to estimate the causal contribution of discussion dynamics.

\vspace{-2mm}
\paragraph{\textbf{Total number of comments}}

\begin{figure}[t]
    \centering
    \subfloat[Total number of comments.]{\includegraphics[width=0.48\linewidth]{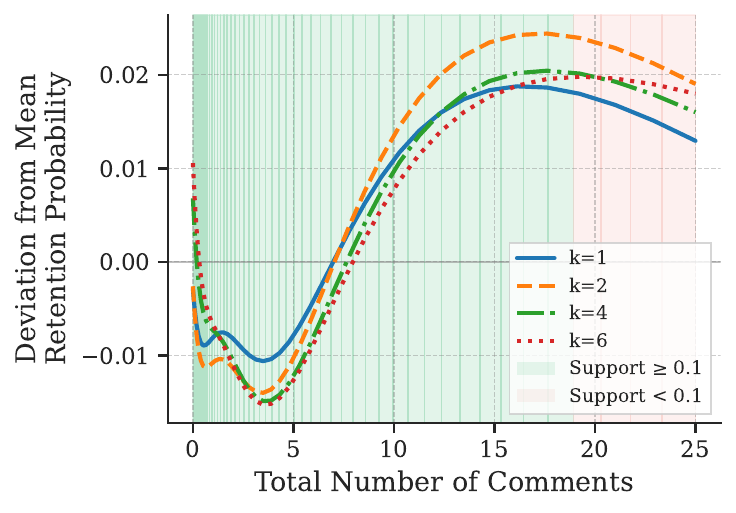}\label{fig:rq3-comments-total}}\hfill
    \subfloat[Maintainer (member) comments.]{\includegraphics[width=0.48\linewidth]{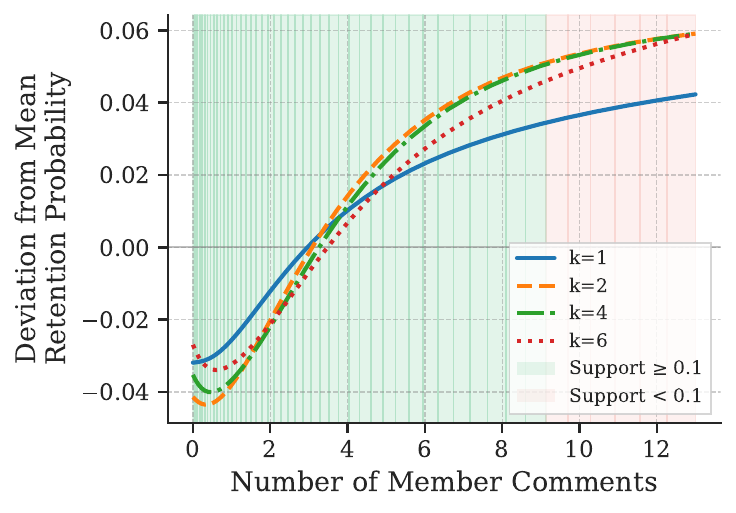}\label{fig:rq3-comments-member}}
    \vspace{-2mm}
    \caption{Estimated causal effects of issue discussion dynamics prior to pull request submission on newcomer retention.}
    \vspace{-4mm}
    \label{fig:rq3-comments}
\end{figure}

We estimate the causal effect of early-stage issue discussion on an issue's intrinsic retention potential, measured as the total number of comments prior to the newcomer's contribution. 
Figure~\ref{fig:rq3-comments}\subref{fig:rq3-comments-total} summarizes the estimated effect of early-stage discussion volume.
In the figure, the curves depict the estimated dose–response curves under different $k$. The x-axis reports the treatment value, while the y-axis reports the estimated retention probability, expressed as the percentage deviation from the average probability in the corresponding data under $k$. A rising curve indicates that increasing the treatment causally improves retention, whereas a declining curve indicates the opposite. 
% 既然已经变成deviation了，这句就不需要了We focus on the shape of the curve (e.g., thresholds, plateaus, and non-linearities) rather than the absolute levels, which can vary with the retained/non-retained definition and modeling choices.
The background shading indicates a \textit{local support ratio}, defined as the fraction of samples whose estimated GPS at treatment level \(t\) exceeds a small threshold \(\varepsilon\) ($\varepsilon=5\%$ in our experiments). This provides a practical diagnostic of the positivity/overlap assumption for continuous treatments~\cite{hirano_PropensityScore_2004} and we therefore interpret regions with support ratio $\ge0.1$ as relatively better supported by the observed data, while treating lower-support regions more cautiously.

The effect can be decomposed into three phases:
First, issues with minimal activity (1--3 comments) exhibit a ``friction dip'', where the retention potential drops to its lowest point. This likely represents scenarios where problems are acknowledged but left ambiguous or unresolved, signaling high risk to potential contributors.
Second, as discussion volume increases (approx. 5--15 comments), we observe a significant ``enrichment climb.'' In this phase, the accumulated comments effectively serve as supplementary documentation, clarifying requirements and reducing uncertainty. This process transforms the issue from a vague problem into a well-scoped, de-risked task, thereby directly elevating its potential to retain a newcomer's effort.
Finally, beyond this optimal range ($>$15 comments), the retention potential begins to decline. This suggests an information overload or complexity trap, where excessive discussion may signal intractability or contentious design disagreements, lowering the issue's attractiveness to newcomers.

% Because these comments are observed before any code contribution, the effect cannot be explained by code review dynamics. Instead, it likely captures the value of pre-implementation coordination and uncertainty reduction.

% \begin{rqsummary}
% \noindent\textbf{Summary.}
% Early-stage discussion causally shapes an issue's retention potential through a non-linear process: initial friction is overcome by a valuable information enrichment phase (5--15 comments), after which excessive discussion signals diminishing solvability.

% \noindent\textbf{Practical Implications.}
% Maintainers should view issue discussion threads as dynamic documentation. Encouraging detailed clarification \emph{before} coding begins is a mechanism to upgrade an issue's quality and de-risk the task for newcomers, but care should be taken to resolve discussions before they become prohibitively long.
% \end{rqsummary}
\vspace{-2mm}
\paragraph{\textbf{Maintainer involvement}}
Figure~\ref{fig:rq3-comments}\subref{fig:rq3-comments-member} illustrates the causal effect of maintainer participation.

Among all factors, maintainer comments exhibit a stronger positive effect on retention potential.
The effect is mostly monotonic: every additional unit of maintainer engagement increases the likelihood that the issue will retain a contributor.
From a potential perspective, the participation of maintainers serves as a high-fidelity signal of task validity and project alignment. It reassures potential contributors that the issue is relevant, the direction is correct, and their future PR is likely to be reviewed. This ``validation effect'' significantly lowers the perceived risk of wasted effort, thereby elevating the issue's intrinsic retention potential.

% \begin{rqsummary}
% \noindent\textbf{Summary.}
% Maintainer involvement is the single strongest predictor of retention potential, acting as a crucial signal of task validity that de-risks the contribution process.

% \noindent\textbf{Practical Implications.}
% Maintainer attention functions as a ``stamp of approval'' that validates an issue's worth. Even lightweight interventions—such as confirming a bug or approving a direction—can significantly boost an issue's attractiveness and retention power.
% \end{rqsummary}
\vspace{-2mm}
\paragraph{\textbf{Sentiment of comments}}
% 这块我把数据集按照Sentiment排序 + 留存label筛选，去找case，但是看下来其实找不到一眼就能够非常直观看到这个结论的case。没有非常典型、dramatic的case适合放在这里
% 或者这样说，直接看issue的case study在retention的场景下本来就不好做，因为看issue本身看它适不适合新手是能够从直觉判断的，但是进一步要看它是不是一个能够留住新手的case，本来就没有那么直观的判断方法。这时我按照sentiment排序找出评论情感指标非常低的留存样本（并且确保后续贡献比较多），最后发现样本之间大差不差，没有非常典型的case可以拿来讲
% We additionally examine the causal effect of comment sentiment on an issue's intrinsic retention potential. Figure~\ref{fig:rq3-comment-sentiment} illustrates the effect.

% Counterintuitively, issues dominated by highly positive sentiment exhibit the lowest retention potential, whereas neutral or mildly negative sentiment correlates with the highest likelihood of newcomer retention. This pattern persists even after controlling for discussion volume and maintainer participation.

\begin{figure}[t]

    \centering
    \includegraphics[width=0.6\linewidth]{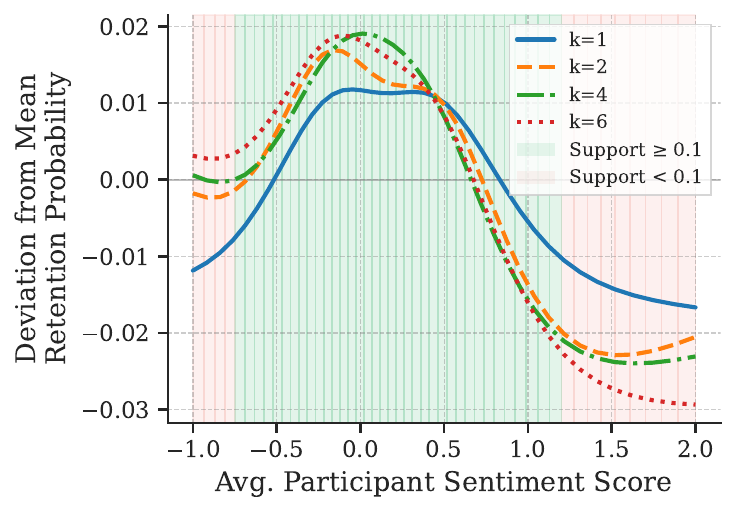}
    \vspace{-2mm} \caption{Estimated causal effect of early-stage comment sentiment on newcomer retention.}
    \label{fig:rq3-comment-sentiment}
\end{figure}

% This result highlights a critical trade-off between emotional validation and information density. Highly positive comments (e.g., brief encouragement or gratitude) often lack substantive technical content, leaving the task ambiguity unresolved. In contrast, neutral or mildly critical discussions are typically rich in objective diagnosis and actionable guidance (e.g., reproduction steps, error logs). This suggests that for an issue to be ``retainable,'' the clarity provided by rigorous technical discourse outweighs the comfort of a friendly but superficial atmosphere.
We further estimate the causal effect of comment sentiment on an issue's intrinsic retention potential. As shown in Figure~\ref{fig:rq3-comment-sentiment}, the relationship is distinctly non-linear. Retention probability is highest for neutral discussions, but lower for strongly positive and very negative ones.

The pattern suggests that newcomer-retaining issues benefit more from a moderate discussion tone. One plausible explanation is that neutral exchanges strike a better balance between social warmth and technical substance, which remain approachable while still conveying concrete diagnostic information, whereas strongly positive comments are often less actionable and strongly negative discussions may discourage participation.

% \begin{rqsummary}
% \noindent\textbf{Summary.}
% Technically grounded, neutral-sentiment discussions causally enhance an issue's solvability and retention potential, outperforming highly positive but low-substance feedback.

% \noindent\textbf{Practical Implications.}
% While a welcoming tone is valuable, it must not substitute for technical substance. Maintainers should prioritize actionable information over purely emotional encouragement to effectively de-risk issues for newcomers.
% \end{rqsummary}

\subsubsection{Reporter Experience}

\begin{figure}[t]
    \centering
    \subfloat[Prior commits.]{\includegraphics[width=0.48\linewidth]{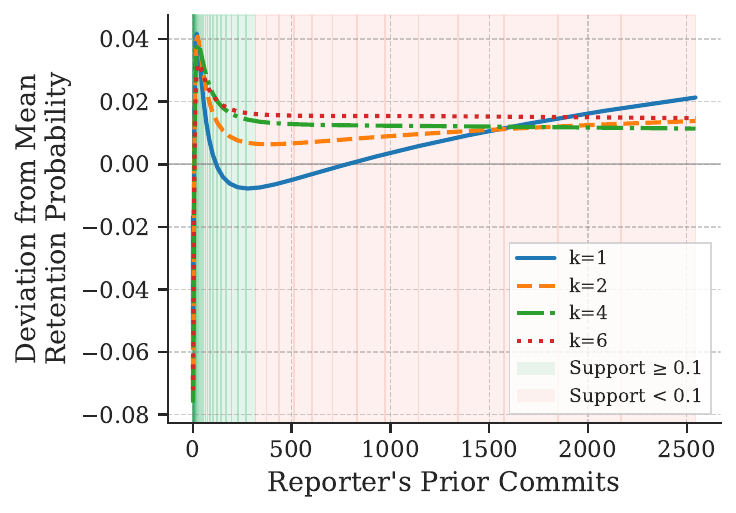}\label{fig:rq3-reporter-exp-commit}}\hfill
    \subfloat[Prior pull requests.]{\includegraphics[width=0.48\linewidth]{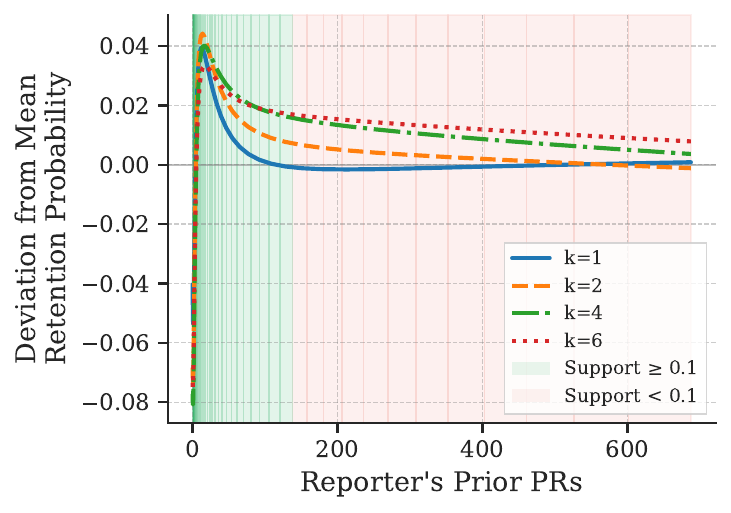}\label{fig:rq3-reporter-exp-pr}}
    \caption{Estimated causal effects of issue reporter experience on newcomer retention, using two activity proxies.}
    \label{fig:rq3-reporter-exp}
\end{figure}

We next examine the experience of the issue reporter within the repository using two complementary proxies (prior commits and prior pull requests). Figure~\ref{fig:rq3-reporter-exp}\subref{fig:rq3-reporter-exp-commit} and Figure~\ref{fig:rq3-reporter-exp}\subref{fig:rq3-reporter-exp-pr} report the corresponding estimated effects.

% The estimated causal effect follows an inverted U-shape: issues reported by moderately experienced contributors yield the highest retention potential, whereas issues reported by complete newcomers or highly experienced contributors are less conducive to retention.

% We conjecture that moderately experienced reporters strike a balance between project-context awareness and a newcomer-like perspective, resulting in clearer and more accessible issue descriptions.
Specifically, issues reported by contributors with moderate experience yield the highest retention potential. In contrast, both complete newcomers and highly experienced maintainers tend to author issues that are less conducive to retention.

This non-monotonic pattern highlights a trade-off in contextual accessibility. While novice reporters often lack the project knowledge to frame actionable tasks (resulting in ambiguity), highly experienced reporters may inadvertently omit critical context due to implicit knowledge, creating a high barrier to entry. Intermediate contributors, possessing both sufficient context and recent onboarding memory, strike an optimal balance, crafting issue descriptions that are technically precise yet accessible to newcomers.

% \begin{rqsummary}
% \noindent\textbf{Summary.}
% Reporter experience exhibits a non-monotonic causal effect on retention potential: intermediate contributors maximize issue accessibility by balancing technical context with clarity, avoiding both the ambiguity of novices and the implicit assumptions of experts.

% \noindent\textbf{Practical Implications.}
% Projects should leverage intermediate contributors as key ``context translators'' in issue triage. Encouraging them to refine or reproduce issues filed by experts can significantly lower the entry barrier for newcomers.
% \end{rqsummary}

\subsubsection{Repository-Level Characteristics}

Our interpretability analysis suggests that several repository-level features are strongly associated with newcomer retention.
However, estimating causal effect for such features is substantially more difficult in our issue-level dataset, because suitable covariates at the repository level are limited. As a fallback, we attempt to examine the repository-level causal effect of \textit{the repository age} and \textit{popularity (\#stars)} on the repository's historical retention ratio, after controlling for contemporaneous repository state variables as covariates.

Nevertheless, post-weighting balance remains inadequate for these covariates according to the Love plots, indicating that residual confounding could not be ruled out under this setting. We therefore do not report the corresponding DRF estimates and interpret these repository-level features as predictive correlates rather than causally identified drivers in our study.

\subsubsection{Sensitivity to Unobserved Confounding}
\label{sec:sensitivity}
We conduct an
E-value sensitivity analysis~\cite{vanderweele_SensitivityAnalysis_2017},
which quantifies the minimum strength of association (on the risk
ratio scale) that an unmeasured confounder would need with both treatment
and outcome to fully explain away the observed effect.

We estimate risk ratios (RR) using modified Poisson regression with
robust (HC3) standard errors~\cite{zou_ModifiedPoisson_2004}, and compute
E-values for both point estimates and confidence interval limits.
For treatments with non-linear dose--response patterns, we additionally
report segment-wise E-values over quantile contrasts.

Overall, effects of reporter experience and maintainer
comments exhibit moderate-to-strong robustness, with E-values up to
2.24, indicating that relatively strong unmeasured confounding would
be required to nullify these effects. In contrast, non-linear factors
such as discussion volume and sentiment yield smaller E-values
(typically around 1.1--1.4). This is expected, as E-values are derived from a single summary effect estimate, whereas the underlying relationships are non-linear. Opposing effects across different ranges
may partially cancel out in the aggregated risk ratio, leading to conservative E-values.
Importantly, these weaker E-values are consistent with our DRF
findings, which reveal non-linear and non-monotonic relationships.

Together, these results suggest that the estimated effects are generally modest and context-dependent, and newcomer
retention is likely driven by multiple interacting factors rather
than a single dominant cause. Accordingly, we interpret findings with lower E-values with appropriate caution and avoid making strong claims about their practical impact. 

\begin{rqsummary}
    \textbf{Summary for RQ3}.
    % The most significant observation is that more positive comment sentiment does not monotonically improve newcomer retention potential. Instead, the highest retention is observed when discussions remain neutral.
    % Besides, \textit{moderately high discussion intensity} (via comments), \textit{participation of project members in comments} and \textit{medium-experienced issue reporter} also improves the retention potential.
    Our causal analysis most consistently suggests that reporter experience and project member participation in issue discussions are beneficial for newcomer retention. For comment sentiment and discussion intensity, the estimated effects are non-linear: retention is highest when discussions are relatively neutral, and moderate levels of discussion appear more favorable than sparse or highly intensive interaction.
\end{rqsummary}

\section{Discussion}
\label{sec:threats}

\subsection{Practical Implications}

\begin{table*}[t]
\centering
\caption{Positioning of key insights relative to prior work. ``Confirmation'' denotes previously reported findings that we provide further evidence; ``New Finding'' denotes patterns not previously documented.}
\label{tab:insights}
\footnotesize % 可考虑换成footnotesize极限压缩空间
\setlength{\tabcolsep}{4pt}
\begin{tabularx}{\textwidth}{@{} p{4.5cm} l p{3.5cm} X @{}}
\toprule
\textbf{Finding} & \textbf{Type} & \textbf{Prior Work} & \textbf{Our Contribution} \\
\midrule

Moderate discussion volume maximizes newcomer retention. & New Finding & \centering --- & Identifies a consistent non-linear causal effect pattern across different retention thresholds $k$. \\
\addlinespace[0.3em]

Maintainer comments exhibit a strong positive effect on retention. & Causal confirmation & Tan et al.~\cite{tan_ItEnough_2023}; Zhou \& Mockus~\cite{zhou_WhoWill_2015} & Provides the first rigorous causal estimation and validation for this phenomenon. \\
\addlinespace[0.3em]

Interaction features outweigh issue-intrinsic attributes. & Confirmation & Steinmacher et al.~\cite{steinmacher_OvercomingOpen_2016} & Quantifies this insight via interpretability analysis on large-scale data. \\
\addlinespace[0.3em]

Neutral discussion tones outperform highly positive ones for retention. & New Finding & \centering --- & Contradicts the intuitive assumption that higher positivity always yields better retention. \\
\addlinespace[0.3em]

Moderately experienced reporters boost newcomer retention. & New Finding & \centering --- & Uncovers an inverted-U effect of reporter experience on newcomer retention. \\

\bottomrule
\end{tabularx}
\end{table*}

Our findings suggest several practical considerations for OSS communities seeking to better support newcomer retention through issue design and interaction practices.
To clearly position our contributions, Table~\ref{tab:insights} outlines these results, separating novel findings from confirmations of existing literature.

First, early discussion dynamics around an issue appears to be closely associated with whether the issue becomes a viable entry point for newcomers. Our adjusted estimates suggest a non-linear pattern, in which a moderate amount of early discussion is associated with higher downstream retention. 
Given the relatively small E-values for this factor, the finding should be interpreted cautiously, although the non-linear dose--response pattern suggests that the sensitivity estimate may be conservative and still offers a potentially useful practical signal.
% One possible interpretation is that early discussion can function as a form of collaborative documentation by surfacing clarifications, constraints, and contextual information before implementation begins. 
In practice, maintainers may benefit from encouraging clarifying questions and technical context early in the discussion, while also helping the conversation converge toward a clear and actionable task definition. We also observe a strong positive trend for project-member participation in issue discussions, for which the E-value analysis indicates moderate robustness to unmeasured confounding. Maintainer attention may serve as a signal that the issue is valid, relevant, and likely to receive follow-up, thereby making the newcomer’s initial contribution experience less ambiguous and more likely to support continued engagement.

Second, highly positive sentiment in issue discussions should not automatically be treated as a reliable signal of a retention-supportive environment. In our data, neutral comment sentiment is estimated to be causally associated with higher retention potential rather than highly positive discussion. This counterintuitive result is noteworthy, but given the modest E-value, it should be treated as a cautious practical signal rather than a definitive recommendation.
% , although this finding should be interpreted with caution because sentiment in technical discussions may be measured imperfectly. 
% One possible explanation is that brief encouragement or gratitude, while socially positive, may sometimes provide less task-relevant guidance than technically specific discussion. 
In practice, maintainers and contributors may benefit more from providing technically actionable feedback than from offering purely affective responses. Information such as reproduction steps, clarification of expected behavior, and concrete diagnostic suggestions can help make newcomers’ first interactions more structured, comprehensible, and rewarding. Specific and technically grounded neutral comments may contribute to newcomer retention by reducing uncertainty, supporting learning, and fostering a sense of productive engagement with the project.
% In practice, maintainers and contributors may wish to prioritize actionable information, such as reproduction steps, clarification of expected behavior, or diagnostic suggestions, over purely affective responses. Neutral or mildly critical discussion, when technically grounded, may sometimes do more to reduce ambiguity and make an issue easier for newcomers to act on.

Third, issues authored by contributors with moderate prior experience appear to be associated with more favorable downstream retention outcomes. A possible interpretation is that such contributors may combine sufficient project familiarity with a relatively recent memory of the onboarding process, allowing them to frame issues in ways that remain understandable to newcomers. From a practical perspective, projects may benefit from paying closer attention to how issue context is constructed during reporting and triage. Encouraging moderately experienced contributors to refine or clarify issue descriptions may help make potential entry points more accessible.

Taken together, these observations suggest that newcomer retention may depend less on identifying inherently “easy” tasks than on shaping issue discussions and surrounding context in ways that reduce uncertainty, clarify expectations, and improve task accessibility.

\subsection{Threats to Validity}
\subsubsection{Internal Validity}
The modeling of an issue is comprehensive and complicated, and our feature extraction may still miss some important indicators to characterize an issue despite a relatively substantial set of features, hence resulting in reduced prediction performance. More comprehensive feature engineering may compensate for this.
In addition, although our causal analysis controls for a rich set of covariates to eliminate systematic bias, unobserved confounding factors may exist and undermine the rigor of causal analysis. This can be alleviated through better feature extraction, but cannot be fully eliminated because observational studies inherently lack control over all potential confounders, and certain latent factors may remain unobserved in GitHub data. We conduct an E-value sensitivity analysis as an addition, which suggests less conclusive evidence for the non-linear effects, whose smaller E-values may partly reflect the limitations of RR-based sensitivity analysis for direction-changing dose--response patterns. Therefore, while the overall conclusions are reasonable, the weaker non-linear findings should be interpreted with caution and further examined in future work.
% Nevertheless, our analysis incorporates the majority of features to reduce the likelihood of severe omitted-variable bias.

\vspace{-1mm}
\subsubsection{External Validity}
According to Section~\ref{subsec:repoc}, although the repositories used in our study are subjected to several quality constraints to ensure their activeness and maintenance, the resulting sample may not fully represent the broader ecosystem of OSS projects on GitHub. Projects with different scales, activity levels, or community structures may exhibit different newcomer retention dynamics.
Besides, our analysis assumes an issue-driven development workflow, where issues play a central role in coordinating tasks and discussions. However, some repositories rely less heavily on issue tracking or use alternative coordination mechanisms (e.g., external communication channels). Such differences in project practices may limit the applicability of our findings. 
Future work could expand the dataset and explicitly distinguish workflow styles to further validate the generalizability of our results.

% \vspace{-1mm}
\section{Conclusion}
\label{sec:conclusion}
This paper presents a large-scale empirical study investigating which types of first issues are most likely to promote newcomer retention in open-source projects. By combining predictive modeling with interpretability analysis and causal inference, we not only assessed the potential of retention prediction models but also revealed fine-grained feature effects on retention outcomes. Our findings highlight actionable patterns that maintainers could leverage to encourage sustained engagement, offering both practical guidance and a methodological template for future studies. Looking forward, these insights can inform retention-oriented issue recommendation systems and further exploration of contextual factors that influence newcomer participation.

\bibliographystyle{ACM-Reference-Format}
\bibliography{retention-ref}

\end{document}